\documentclass[twocolumn, prl, superscriptaddress]{revtex4-1}
\usepackage{amssymb}
\usepackage{amsmath}
\usepackage{epsfig}
\usepackage{graphics, graphicx}
\usepackage{bbold}
\usepackage{psfrag}
\usepackage{mathcomp}
\usepackage{subfigure}
\usepackage{verbatim}
\usepackage{bm}
\usepackage[marginal]{footmisc} 
\usepackage{color}
\usepackage{hyperref}
\def\cp#1{\mathbf{#1}}

\begin{document}

\title{Universal Bound States with Bose-Fermi Duality in Microwave-Shielded Ultracold Molecules}
\author{Tingting Shi}\thanks{These authors contributed equally to this work.}
\affiliation{Beijing National Laboratory for Condensed Matter Physics, Institute of Physics, Chinese Academy of Sciences, Beijing, 100190, China}
\author{Haitian Wang}\thanks{These authors contributed equally to this work.}
\affiliation{Beijing National Laboratory for Condensed Matter Physics, Institute of Physics, Chinese Academy of Sciences, Beijing, 100190, China}
\affiliation{School of Physical Sciences, University of Chinese Academy of Sciences, Beijing 100049, China}
\author{Xiaoling Cui}
\email{xlcui@iphy.ac.cn}
\affiliation{Beijing National Laboratory for Condensed Matter Physics, Institute of Physics, Chinese Academy of Sciences, Beijing, 100190, China}
\date{\today}

\begin{abstract}
We report universal bound states of  microwave-shielded ultracold molecules that solely depend on the strengths of long-range dipolar interaction and microwave coupling. Under a highly elliptic  microwave field, few-molecule scatterings in three dimension are shown to be governed by effective one-dimensional (1D) models, which  well reproduce the tetratomic bound state and the Born-Oppenheimer potential in three-molecule sector. For hexatomic systems comprising three identical molecules, we find much deeper bound state than the tetratomic one, with binding energy exceeding twice of the latter. Strikingly,  these bound states display Bose-Fermi duality as facilitated by the effective 1D scattering with a large repulsive core from angular fluctuations.  For large molecule ensembles, our results suggest the formation of elongated self-bound droplets with crystalline patterns in both bosonic and fermionic molecules.

\end{abstract}
\maketitle

As an ideal platform for quantum simulation with strong long-range interactions, ultracold polar molecules have recently achieved great developments due to the application of microwave shielding\cite{Hutson,Quemener,Doyle,Luo1,Luo2,Luo3,Wang,Will1,Will2,Will_new,Wang2}. With this technique, inter-molecule potential shows, apart from an anisotropic long-range tail, a large repulsive shielding core (spanning hundreds to thousands of Bohr radii) that efficiently suppresses two-body  losses. This facilitates the realizations of Fermi degenerate gas of NaK molecules\cite{Luo1} and Bose-Einstein condensation of NaCs\cite{Will2,Will_new} and NaRb\cite{Wang2} molecules. Further tuning the microwave ellipticity enhances inter-molecule attraction, leading to  scattering resonance\cite{Luo2} and field-linked tetratomic molecules\cite{Luo3} observed in NaK molecular gas. Theoretically, interesting many-body phases of self-bound bosonic droplets\cite{Shi1,Shi2,Langen1,Langen2} and pairing fermion superfluids\cite{Shi3}  have been revealed, with the former recently explored in experiments\cite{Will_new,Wang2}. However, in the fundamental few-body level, despite significant progress in two-body (or two-molecule) properties under  shielding potentials\cite{Bohn, Bohn2, Shi3,Croft,Hutson2,Zhang,Hutson4}, intriguing few-body phenomena beyond two-body ones remain largely unexplored\cite{Greene, DWWang}.

\begin{figure}[t]
\includegraphics[width=8cm]{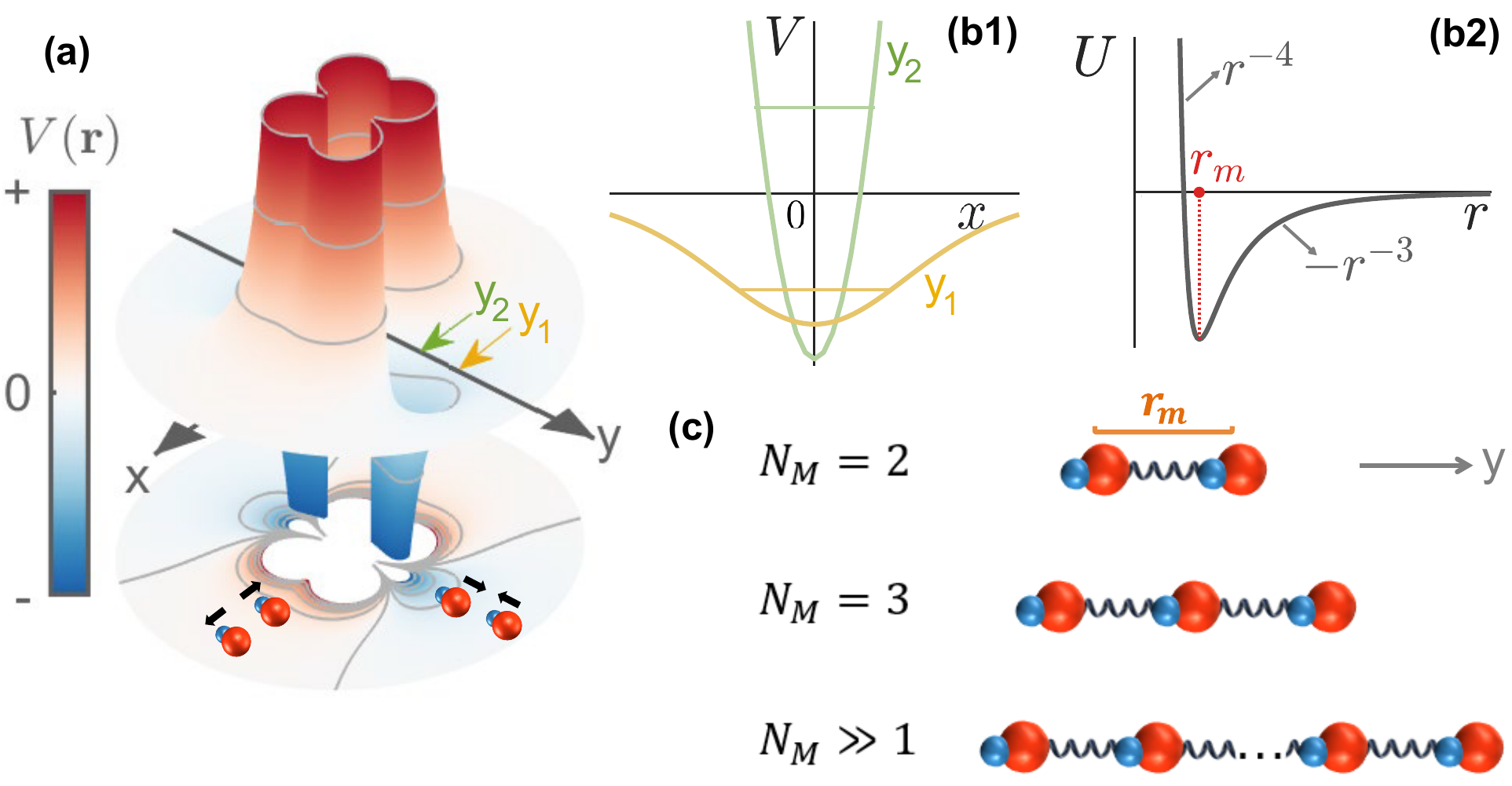}
\caption{(Color Online). Schematics of interaction potentials and bound states of polar molecules  shielded by a highly elliptic microwave field. (a) Interaction potential $V({\cp r})$ at $xy$ plane ($z=0$). (b1) Slices of $V$ at different $y$ ($y_1>y_2$, as marked by arrows in (a)). Fluctuations along $x$ ($\sim\delta\phi$) lead to a zero-point energy and effectively upshift the potential to horizontal level. The resulted effective potential $U(r\equiv |y|)$ is plotted in (b2), showing a long-range attraction $(-r^{-3})$ and a repulsive core $(r^{-4})$ from angular fluctuations. Its minimum is located at $r_m$. (c) Ground state distribution of bosonic or fermionic molecules with number $N_M$. They are all bound states aligned along $y$ with typical inter-molecule distance $r_m$.  
}  \label{fig_illustration}
\end{figure}

Few-body physics has been extensively studied in ultracold atomic systems with short-range interactions\cite{review_PR,review_RMP,review_RPP}, where many fascinating cluster bound states were discovered. For instance, Efimov states\cite{Efimov1,Efimov2}, characterized by discrete scaling symmetry and energy sensitivity to short-range parameters, dominate in identical bosons, three distinguishable particles and highly mass-imbalanced fermion mixtures, often driving atom losses. In contrast, universal bound states, irrelevant to short-range details and stable against inelastic collision, exist in fermion mixtures with intermediate mass imbalance\cite{KM, Blume, Petrov,  Pricoupenko, Parish,Cui, KM_1D,Mehta_1D,Petrov_1D} and have been shown to induce novel quantum phases with high-order correlations\cite{Parish3, Parish4, Naidon, mass_polaron, QSF}. How these distinct bound states behave in long-range interacting polar molecules is an interesting yet challenging problem. Along this direction, previous theories have investigated Efimov physics for three particles interacting with anisotropic long-range and isotropic short-range potentials\cite{Greene1,Greene2,Endo}. Such potentials, however, are substantially different from those in microwave-shielded polar molecules. In particular, a large dipole length and a large repulsive shielding core in the latter case may  support a new type of universal clusters --- a possibility that demands rigorous exploration.

In this work, we present the first theoretical investigation of hexatomic (three-molecule) bound state in microwave-shielded 3D polar molecules. To maximize binding strength,  we consider the microwave field linearly polarized along $y$ (with elliptic angle $\xi=\pi/4$). 
Effective one-dimensional (1D) models are established for few-molecule scattering in 3D, with effective potentials  containing a long-range $-r^{-3}$ attraction  and a $r^{-4}$ repulsive core  from angular fluctuations, see illustration in Fig.\ref{fig_illustration}. These models well reproduce  tetratomic bound states as well as the Born-Oppenheimer potential in three-molecule sector. Applying to three identical molecules, we find hexatomic bound state is more favored than tetratomic state, with binding energy exceeding twice of the latter. All these states are universal, in that their properties only depend on physical parameters of long-range dipolar strength and microwave coupling.
Importantly,  effective 1D scattering with a large repulsive core also facilitates the Bose-Fermi duality of these bound states, i.e., their energies and spatial densities are identical between bosonic  and fermionic systems.  Extending to large ensembles, our results suggest the formation of elongated self-bound droplets with crystalline pattern in both bosonic  and fermionic molecules.

We adopt the interaction potential of microwave-shielded molecules\cite{Shi3} and extrapolate it to elliptic angle $\xi=\pi/4$, where the microwave field ${\cp E}= Ee^{i(kz-\omega t)}({\cp e}_+\cos\xi  +{\cp e}_-\sin\xi)+c.c.$ (with ${\cp e}_{\pm}=\mp ({\cp e}_x\pm i{\cp e}_y)$) is linearly polarized along $y$ and
\begin{eqnarray}
V({\cp r})&=&\frac{C_3}{r^3} \Big(
	3\cos^2\theta -1 +3 \sin^2\theta\cos(2\phi)\Big) \nonumber\\
	&&+ \frac{C_6}{r^6} \Big(\sin^2\theta\sin^2(2\phi) + \sin^2(2\theta)\sin^4\phi \Big). \label{V3d}
\end{eqnarray}
Here ${\cp r}=(r,\theta,\phi)$ is the inter-molecule distance; $C_3=\frac{d^2}{48\pi\epsilon_0(1+\delta_r^2)}$,  $C_6 = \frac{d^4}{128\pi^2\epsilon_0^2\Omega(1+\delta_r^2)^{3/2}}$ ($\delta_r = \frac{|\delta|}{\Omega}$) with $\Omega$ and $\delta$, respectively, denoting the coupling and detuning of microwave field; $d$ is dipole moment that defines dipole length $l_d\equiv \frac{m}{\hbar^2}\frac{d^2}{48\pi\epsilon_0}$. NaK\cite{Luo1,Luo2,Luo3}, NaRb\cite{Wang}, NaCs\cite{Will1,Will2} molecules all have large dipole lengths $l_d/(10^4a_0)\approx 1.1,\ 2.6,\ 8.3$ ($a_0$ is Bohr radius).  In this work, we assume a small $\delta_r=0.2$, and  take the length and energy units as $l_u=l_d/20$ and $E_u=\frac{\hbar^2}{ml_u^2}$. For NaK, this gives  $l_u\approx 550a_0$, and $\hbar\Omega/E_u \approx 52$ at a typical $\Omega=(2\pi)10$MHz. 

As shown in Fig.\ref{fig_illustration}(a), $V({\cp r})$ is extremely anisotropic: it is fully attractive ($=-4C_3/r^3$) along $y$ while fully repulsive  ($=2C_3/r^3$) along $x$ (or $z$), and the repulsion $\sim C_6/r^6$ takes place in general directions except $x,y,z$. Among all $\xi$, the present case ($\xi=\pi/4$) has the most pronounced attraction along $y$ and thus most favors bound state formation, as also inferred from recent NaK experiment\cite{Luo3}. 

\begin{figure}[t]
\includegraphics[width=7cm]{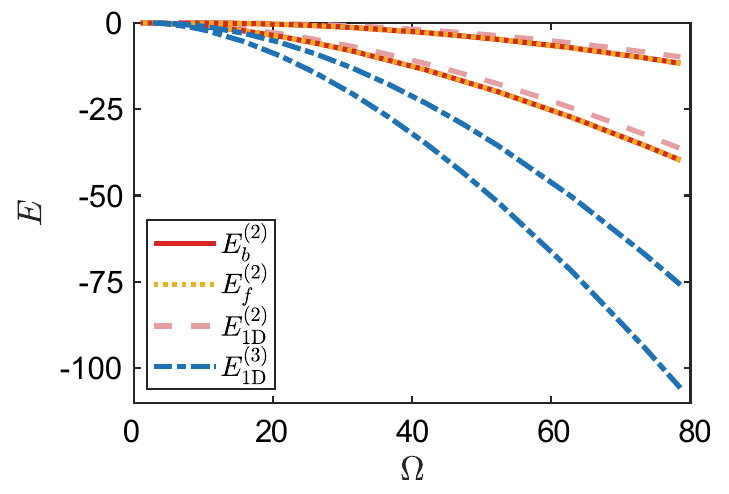}
\caption{(Color Online). Binding energies of two and three identical molecules as functions of microwave coupling $\Omega$. Red solid and yellow dot lines show exact tetratomic energies of bosonic ($E_b^{(2)}$) and fermionic ($E_f^{(2)}$) systems,   in comparison to pink dashed lines from effective 1D model ($E^{(2)}_{\rm 1D}$). Blue dash-dot lines show hexatomic energies from effective 1D model ($E^{(3)}_{\rm 1D}$). For each case we show two lowest energy levels.  
The units of $E$ and $\Omega$ are $E_u$ and $E_u/\hbar$.}  \label{fig_E}
\end{figure}

\begin{figure}[t]
\includegraphics[width=8.5cm]{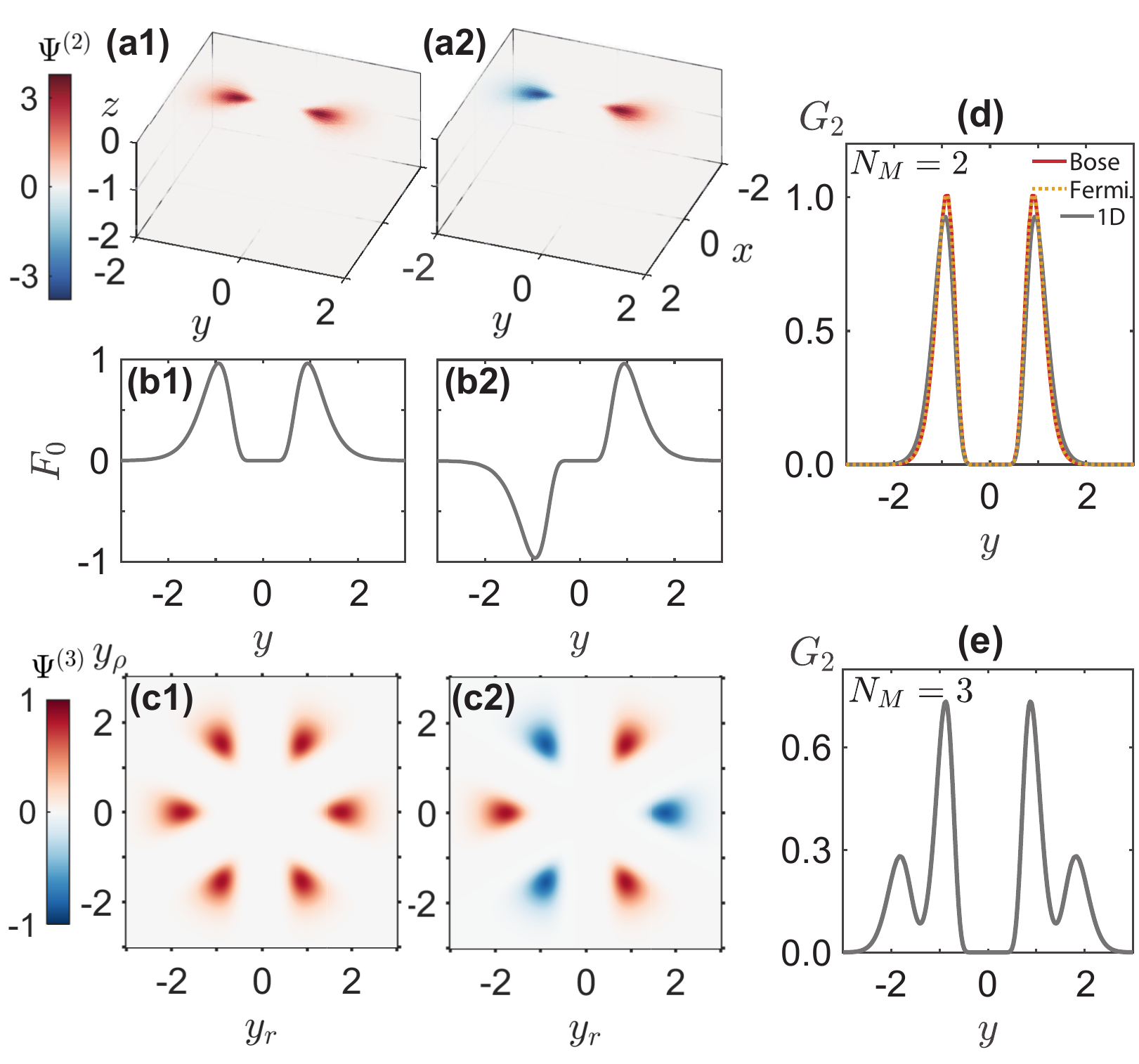}
\caption{(Color Online). Bose-Fermi duality of tetratomic and hexatomic bound states at $\hbar\Omega/E_u=52$. (a1,a2) and (b1,b2) show wavefunctions of the lowest tetratomic states, respectively, from exact solution and effective 1D model. (c1,c2) are wavefunctions of the lowest hexatomic states from effective 1D model. (a1,b1,c1) are for bosonic molecules and (a2,b2,c2) are for fermionic ones.  (d) and (e) are density correlation functions $G_2(y)$  for tetratomic and hexatomic states. In (d), exact results of bosonic (red solid) and fermionic (yellow dot) systems are compared with those from effective 1D model (gray solid). Here the length unit is $l_u$.}  \label{fig_duality}
\end{figure}

To begin with, we exactly solve the tetratomic binding energies of bosonic ($E_b^{(2)}$) and fermionic ($E_f^{(2)}$) molecules by diagonalizing their relative Hamiltonian $H^{(2)}=-\hbar^2\nabla_{\cp r}^2/m +V({\cp r})$ in momentum $\{k\}$ and angular momentum $\{lm\}$ space\cite{supple}. Fig.\ref{fig_E} shows $E_{b,f}^{(2)}$ for two lowest bound states, which emerge one by one as increasing $\Omega$. Remarkably, despite of distinct $\{klm\}$ compositions in bosonic and fermionic solutions, $E_{b}^{(2)}$ and $E_{f}^{(2)}$ are almost identical, with only clear deviations near their emergence\cite{supple}.  Their typical wavefunctions $\Psi_{b,f}^{(2)}({\cp r})$ are shown in Fig.\ref{fig_duality}(a1,a2): although they have opposite symmetries in two systems, their absolute values are identical --- both peak at finite $y$ and vanish at ${\cp r}=0$.

To understand above phenomena, we study the effective scattering of two molecules along $y$ direction. Treat any angular deviations from $y$ as small fluctuations:
\begin{equation}
\delta\theta\equiv \theta-\pi/2, \ \ \ \delta\phi\equiv \phi-\phi_0, 
\end{equation}
with $\phi_0=\pm\pi/2$, we can expand $V({\cp r})$ up to the lowest fluctuation order $\sim \delta\theta^2, \delta\phi^2$.  
Together with the kinetic term, 
 $H^{(2)}$ then reduces to two independent harmonic oscillators with respect to $\delta\theta$ and $\delta\phi$\cite{supple}. 
Further following the adiabatic representation\cite{Blume2,Cavagnero,Greene3}, we write the tetratomic state as
$\Psi^{(2)}({\cp r})=\frac{1}{r}\sum_{\nu} F_{\nu}(y) \psi_{\nu}(y;\delta\theta,\delta\phi)$, with $y=r\sin\phi_0$ the reduced coordinate  and 
$\psi_{\nu}$  the $\nu$-th eigen-state of harmonic oscillators.  
Neglecting the off-diagonal couplings between different $\nu$, we obtain the reduced 1D equation for the ground state ($\nu=0$):
\begin{equation}
\left( - \frac{\hbar^2}{m} \frac{\partial^2}{\partial y^2} + U^{(2)}(|y|) \right) F_{0}(y) = E^{(2)}_{\rm 1D}F_{0}(y), \label{Heff_2}
\end{equation}
with ($r\equiv|y|$) 
\begin{equation}
U^{(2)}(r)=-\frac{4C_3}{r^3}+\sqrt{\frac{4\hbar^2}{m}\left(\frac{6C_3}{r^5} + \frac{4C_6}{r^8}\right)}. \label{Ueff2}
\end{equation} 
In this way, $F_{0}$ and $U^{(2)}$ can be viewed as the reduced wavefunction and potential in effective 1D. $U^{(2)}$ features a long-range attraction $\sim -r^{-3}$ and a repulsive core $\sim r^{-4}$ that stems from zero-point energy of angular fluctuations (Fig.\ref{fig_illustration}(b1,b2)). Physically, such $r^{-4}$ repulsion originates from the interplay of kinetic motion and $C_6/r^6$ shielding potential, which is very robust and applicable to general $\xi$. Clearly, this repulsion dominates at small $r$  and effectively forbids two molecules coming close, as evidenced by the vanishing of $F_{0}(y=0)$ and $\Psi^{(2)}({\cp r}=0)$ in Fig.\ref{fig_duality}(a1,a2,b1,b2), regardless of the statistics of these molecules.  Therefore, the bosonic and fermionic systems share identical energies and spatial densities, while the statistics just determine the symmetry of their wavefunctions. 
This can be viewed as an extension of Bose-Fermi duality from 1D short-range interacting  systems\cite{Girardeau1,Cheon} to 3D polar molecules with anisotropic long-range interaction.  Similar phenomenon has also been indicated for Efimov trimers in dipolar systems\cite{Endo}. 

As shown in Fig.\ref{fig_E}, $E_{\rm 1D}^{(2)}$ produced by Eq.(\ref{Heff_2}) match  well with exact $E_{b,f}^{(2)}$ across a wide range of $\hbar\Omega/E_u$. The Bose-Fermi duality can be probed through the density correlation function along $y$:
\begin{equation}
G_2(y)\equiv \langle n(0)n(y)\rangle, \label{G2}
\end{equation}
which follows $|F_{0}(y)|^2$ from effective 1D model and $\int dx dz |\Psi_{b,f}^{(2)}({\cp r})|^2$ from exact 3D result. Fig.\ref{fig_duality}(d) shows that these $G_2$ agree excellently with each other. In fact, the 1D and 3D wavefunctions match very well over a wide range of $\Omega$, with overlap $>97\%$ for $\hbar\Omega/E_u>10$\cite{supple}. These agreements demonstrate the validity of effective 1D treatment in tetratomic problem. 

Now we come to hexatomic system of three molecules. A key issue here is to understand the effect of a third molecule,  in particular, whether it can bring a scale-invariant $\sim -R^{-2}$ potential as in Efimov physics\cite{review_PR,review_RMP,review_RPP}. A physically transparent way to approach it is from the Born-Oppenheimer (BO) limit, by studying the induced heavy-heavy potential by a light object\cite{BOA}. Importantly, here we can also use BO analysis to benchmark the effective 1D approach to hexatomic problem. 

\begin{figure}[t]
\includegraphics[width=7cm]{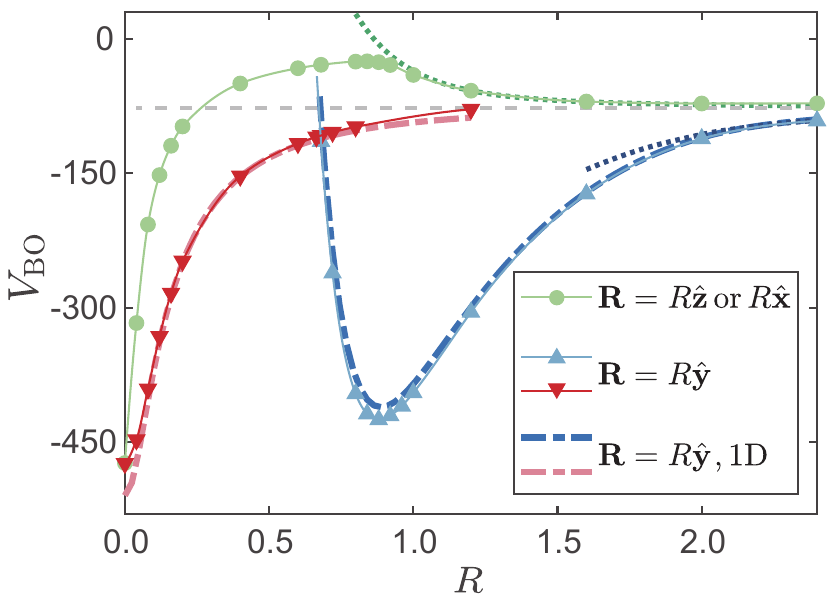}
\caption{(Color Online). Light-induced heavy-heavy potential $V_{\rm BO}(R\equiv|{\cp R}|)$ in the Born-Oppenheimer limit. Here $\hbar\Omega/E_u=52$, and the gray horizontal line marks the binding energy of one heavy-light pair.  For ${\cp R}$ along $y$, we show two orthogonal levels of $V_{\rm BO}$ (blue and red triangles), while for ${\cp R}$ along $x$ or $z$ (equivalent), we show the lowest  $V_{\rm BO}$ (green circle). Dashed lines are results from effective 1D model (Eq.\ref{U_L}). Dotted lines  at large $R$ show mean-field energies between a heavy-light pair and the rest light molecule\cite{supple}. The length and energy units are respectively $l_u$ and $E_u$.}  \label{fig_BO}
\end{figure}

For a light molecule ($\cp r$) interacting with two heavy ones ($\pm{\cp R}/2$) via microwave-shielded potential, we have exactly diagonalized its Hamiltonian $H_L=-\frac{\hbar^2}{2m}\nabla^2_{\cp r} + V({\cp r}-\frac{\cp R}{2})+V({\cp r}+\frac{\cp R}{2})$\cite{supple} and obtained the eigen-energy 
$V_{\rm BO}({\cp R})$, which is exactly the light-induced heavy-heavy potential.
In Fig.\ref{fig_BO}, we plot out $V_{\rm BO}$ for different orientations of ${\cp R}$ at a typical $\Omega$, where the effective 1D approach works well to tetratomic problem.  We can see  $V_{\rm BO}({\cp R})$ exhibits strong anisotropy inherited from the microwave shielding: it is the lowest for ${\cp R}$ along $y$, while much higher along $x$ or $z$.  Details of  $V_{\rm BO}({\cp R})$  can be found in \cite{supple}.   An important message here is that for all ${\cp R}$, we do not observe $-R^{-2}$ behavior in $V_{\rm BO}$, suggesting the Efimov physics be greatly suppressed in this case.

The BO potential for ${\cp R}$ along $y$ can be well understood from effective 1D theory. Assume small angular fluctuations of ${\cp r}$ from $y$ direction, we can expand $H_L({\cp r})$ up to the lowest  fluctuation order and extract the  zero-point energy. This  contributes to a repulsive force in the effective 1D potential\cite{supple}:
\begin{equation}
U_L(y)=-4C_3\left(\frac{1}{|y_-|^3}+\frac{1}{|y_+|^3} \right)+u_L(y), \label{U_L} 
\end{equation}
where $y=r\sin\phi_0$, $y_{\pm}=y\pm R/2$ and $u_L$ is the induced repulsive force.
At $y_{\pm}\rightarrow 0$, $u_L\sim |y_{\pm}|^{-4}$, signifying a strong repulsion between a heavy-light pair at short distance. This shows the robustness of  $r^{-4}$ repulsive core, as established in both two- and three-molecule scattering sectors.  As shown in Fig.\ref{fig_BO}, $V_{\rm BO}$ from effective 1D theory fit exceedingly well with exact results,  demonstrating the validity of 1D treatment to hexatomic system. 

Finally we  turn to three identical molecules. Similar problem with a shielding potential has been studied in 2D\cite{DWWang}. However, 3D case  is notoriously difficult to solve  due to larger Hilbert space and time-consuming computation of long-range coupling strengths. To overcome these difficulties, we resort to the effective 1D approach, given its success in both tetratomic and mass-imbalanced hexatomic problems. 

\begin{figure}[t]
\includegraphics[width=8cm]{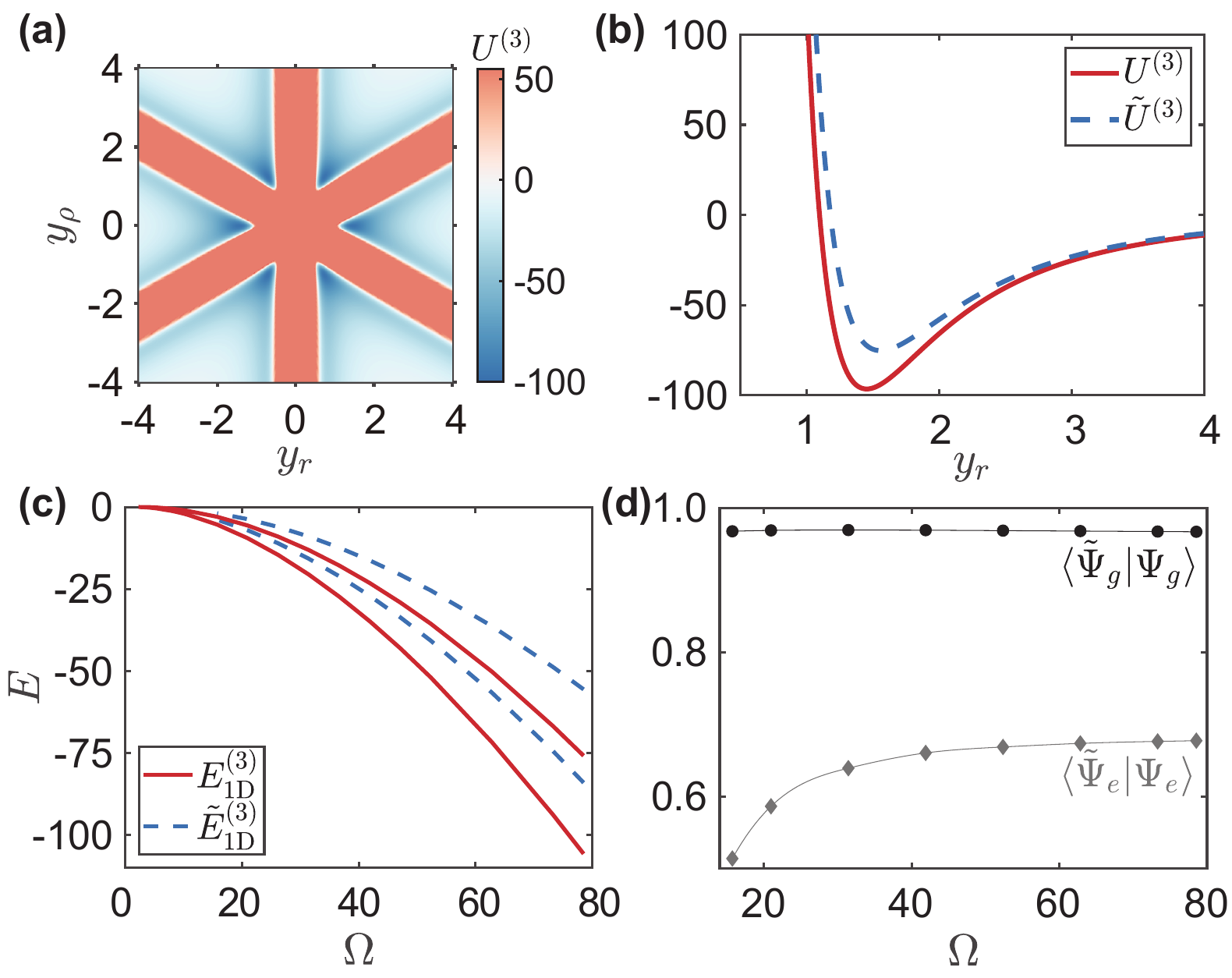}
\caption{(Color Online). Effective potential and wavefunctions of three identical molecules. (a) Effective 1D potential $U^{(3)}$ (Eq.\ref{U0}) in $(y_r,y_{\rho})$ plane at $\hbar\Omega/E_u=52$.  (b)  $U^{(3)}$ and $\tilde{U}^{(3)}$ as functions of $y_r$ at a fixed $y_{\rho}=0$ (corresponding to three equally spaced  molecules).  (c) Energy spectra from $U^{(3)}$ (solid) and $\tilde{U}^{(3)}$ (dashed). (d) Wavefunction overlap for the ground and excited states of two models. The units of length, energy, and $\Omega$ are respectively $l_u$, $E_u$ and $E_u/\hbar$.  }  \label{fig_3body}
\end{figure}

In the center-of-mass frame, three identical molecules (${\cp r}_1,{\cp r}_2,{\cp r}_3$) can be described by two relative coordinates 
 ${\cp r}={\cp r}_2-{\cp r}_1$ and $\bm{\rho}=\frac{2}{\sqrt{3}}({\cp r}_3-({\cp r}_1+{\cp r}_2)/2)$, whose deviations from $y$ direction give four fluctuation variables $\{\delta\theta_{r},\delta\phi_{r}, \delta\theta_{\rho},\delta\phi_{\rho}\}$. The expansion of $H^{(3)}({\cp r},\bm{\rho})$ up to the lowest fluctuation order leads to two sets of coupled harmonic oscillators in terms of $\{\delta\theta_{r},\delta\theta_{\rho}\}$ and $\{\delta\phi_{r}, \delta\phi_{\rho}\}$. 
Their zero-point energies comprise the repulsive force for three molecules effectively moving along $y$, denoted as $u^{(3)}$\cite{supple}. Finally we arrive at the effective 1D model  $H^{(3)}_{\rm 1D}= -\frac{\hbar^2}{m}\left( \frac{\partial^2}{\partial y_{r}^2} + \frac{\partial^2}{\partial y_{\rho}^2}\right) + U^{(3)}(y_r,y_{\rho})$, with
\begin{equation}
U^{(3)}=-4C_3\left(\frac{1}{|y_r|^3}+\frac{1}{|y_{-}|^3}+\frac{1}{|y_{+}|^3} \right)+u^{(3)}(y_r,y_{\rho}), \label{U0}
\end{equation} 
where $y_r= y_2-y_1,\ y_{\rho}= \frac{2}{\sqrt{3}}(y_3-(y_1+y_2)/2)$ are the projections of ${\cp r},\ \bm{\rho}$ along $y$, respectively, and $y_{\pm}=\frac{y_r}{2}\pm\frac{\sqrt{3}{y_{\rho}}}{2}$. Fig.\ref{fig_3body}(a) shows the typical structure of $U^{(3)}(y_r,y_{\rho})$, which fully respects  exchange symmetry under $y_i\leftrightarrow y_j$. 

We have solved hexatomic bound states by exactly diagonalizing $H^{(3)}_{\rm 1D}$ in discretized $(y_r,y_{\rho})$ space. The binding energies of two lowest hexatomic states as functions of $\Omega$ are shown in Fig.\ref{fig_E}. Remarkably, these  states are generally much deeper than tetratomic ones: for the ground state, its binding energy is even beyond twice of tetratomic case. In addition, all these states obey Bose-Fermi duality, as shown by $\Psi^{(3)}(y_r,y_{\rho})$ in Fig.\ref{fig_duality}(c1,c2). This, again, is protected by $\sim r^{-4}$ repulsion in $u^{(3)}$, which prohibits two molecules coming close together. The resulted crystalline pattern  of three molecules can be detected from $G_2(y)$, as shown in Fig.\ref{fig_duality}(e). 

To gain a physical picture for hexatomic binding, we consider a decomposed form of interaction potential:
\begin{equation}
\tilde{U}^{(3)}=U^{(2)}(y_r)+U^{(2)}(y_-)+U^{(2)}(y_+). \label{U0_2}
\end{equation}
$\tilde{U}^{(3)}$ can be a good approximation for $U^{(3)}$ (see Fig.\ref{fig_3body}(b)), and importantly, it inspires us  to construct hexatomic states from tetratomic ones. To be concrete, for three molecules aligning along $y$ with $y_1<y_2<y_3$, the 1D Hamiltonian under $\tilde{U}^{(3)}$ can be written as\cite{supple}  
\begin{equation}
\tilde{H}_{\rm 1D}^{(3)}=H^{(2)}_{\rm 1D}(y_{12})+H^{(2)}_{\rm 1D}(y_{23})+h'(y_{12},y_{23}),  \label{Hdimer}
\end{equation} 
where $y_{ij}\equiv y_{j}-y_{i}$, $H_{\rm 1D}^{(2)}$ is given in (\ref{Heff_2}), and $h'(y,y')=-\frac{\hbar^2}{m}\frac{\partial}{\partial y}\frac{\partial}{\partial y'}+U^{(2)}(y+y')$. In Eq.(\ref{Hdimer}), two $H_{\rm 1D}^{(2)}$ give neighboring  tetratomic states, while $h'$ builds long-range correlation between them. Take two tetratomic levels $\Psi^{(2)}_g$ and $\Psi^{(2)}_e$, we then expand $\tilde{H}^{(3)}$ in the bases $\{\Psi^{(2)}_g(y_{12})\Psi^{(2)}_g(y_{23}),\ \Psi^{(2)}_g(y_{12})\Psi^{(2)}_e(y_{23}),\ \Psi^{(2)}_e(y_{12})\Psi^{(2)}_g(y_{23})\}$.  
The resulted spectra $\tilde{E}_{\rm 1D}^{(3)}$ show qualitative agreements with $E_{\rm 1D}^{(3)}$ in Fig.\ref{fig_3body}(c). The ground state wavefunction, dominated by $\Psi^{(2)}_g(y_{12})\Psi^{(2)}_g(y_{23})$, also matches very well with full 1D result  in Fig.\ref{fig_3body}(d). 
This explains why the lowest hexatomic state coexists  with tetratomic state,  with binding energy beyond twice of the latter due to negative energy correction from $h'$.
For the first excited state, it involves local excitation of one tetratomic state from $\Psi^{(2)}_g$ to $\Psi^{(2)}_e$\cite{supple}, and its overlap with full 1D result is above $60\%$ for $\hbar\Omega/E_u\gtrsim20$, see Fig.\ref{fig_3body}(d).

We emphasize that all above bound states are universal, in that their properties only depend on the long-range dipolar strength (with length scale $\sim l_{d}$) and the microwave coupling (with length scale $l_{\Omega}\equiv \sqrt{\hbar/(m\Omega)}$). That is to say, for different polar molecules with the same $l_d$ and $l_{\Omega}$, they  share the same bound state property.   This is distinct from Efimov trimers in atomic systems, whose spectrum requires the knowledge of a (short-range) three-body parameter that depends on specific atoms\cite{Efimov1,Efimov2, Grimm_expt}.
With dipolar interactions, it has been shown that the Efimov physics emerges at length scale $l_d\ll r\ll |a_s|$ ($a_s$ is s-wave scattering length)\cite{Greene1}. Given polar molecules with very large $l_d\gtrsim 10^4 a_0$\cite{Luo1,Luo2,Luo3, Wang, Will1,Will2}, Efimov trimers are  expected to be very shallow and only appear in a narrow window near shape resonance $a_s\rightarrow\infty$\cite{You, Blume2, Cavagnero, Qi_Zhai}. These make  Efimov physics extremely difficult to explore in practice. On the contrary, our universal clusters have  typical size $\sim r_m (\ll l_d)$, as located by the minimum of $U^{(2)}$ (Eq.\ref{Ueff2}), and thus are much more realistic to probe. 
In the regime $l_{\Omega}\ll l_d$, we have  $r_m\approx4\sqrt{2} l_{\Omega}$ around hundreds of $a_0$ for typical $\Omega\sim$ tens of MHz: for instance, $r_m=428a_0$ for NaK molecules at $\Omega=(2\pi)10$MHz. For these universal clusters (of size $\sim r_m$), the effective 1D approach works well and the Bose-Fermi duality can be guaranteed. They should be distinguished from shallow clusters near resonance, where quantum statistics play a significant role and the clusters carry a 3D instead of 1D character\cite{supple}.

Above analysis of universal clusters can be directly extended to large ensemble of molecules.  For $N$ identical (bosonic or fermionic) molecules, since any neighboring molecules are closely linked along $y$ as tetratomic bound state, the ground state under 1D description would be $\Psi^{(2)}_g(y_{12})\Psi^{(2)}_g(y_{23})...\Psi^{(2)}_g(y_{N-1,N})$ at given order $y_1<y_2...<y_N$. The whole system then behaves as an elongated and crystalline self-bound droplet, see Fig.\ref{fig_illustration}(c). The additional long-range correlation between molecules leads to an attractive energy correction that could further stabilize the system\cite{supple}. The density crystallization can be probed via $G_2(y)$ (Eq.\ref{G2}) using quantum gas microscopes\cite{correlation_ENS, correlation_MIT1,correlation_MIT2}. The ground state droplet can undergo local excitations with $\Psi^{(2)}_g(y_{i,i+1})\rightarrow \Psi^{(2)}_e(y_{i,i+1})$, where $i$ is the index of tetratomic link. By manipulating the locations of $\{i\}$, one could imagine a rich excitation manifold with various crystalline patterns.

Our work has revealed unique few-body physics in polar molecules that are substantially different from those in short-range interacting atomic systems. First, the strong dipolar interaction with a large  shielding repulsion is shown to favor the formation of universal bound states, whose properties only depend on two physical lengths ($l_d$ and $l_{\Omega}$). Secondly, the interaction anisotropy facilitates the effective 1D description and leads to fascinating property of Bose-Fermi duality, despite the physical system is in 3D free space. Finally, the many-body implication of these universal clusters are very different from that in atomic systems. Specifically, clusters in the latter case are often viewed as composite particles in driving collective many-body phases\cite{Parish3, Parish4, Naidon, mass_polaron, QSF}. However, here the spatially extended clusters cannot be considered as composite unit. Instead, when adding more molecules to the cluster, it will become a new bigger bound state and finally evolve  to a self-bound droplet (Fig.\ref{fig_illustration}(c)). 

We expect our results of universal clusters and elongated droplets are generally applicable to finite $\xi(<\pi/4)$, where microwave field is elliptic enough to support 1D scenario. In fact, a smaller $\xi$ can efficiently suppress inelastic two-body loss\cite{Huston5} and make the analytical potential in (\ref{V3d}) quantitatively more accurate\cite{Shi3}.
With a smaller $\xi$, a bare  $r^{-6}$ shielding core  will add to the induced $r^{-4}$ repulsion, leading to shallower bound states. In this case, the effective 1D theory can be further improved  by incorporating high-order angular fluctuations\cite{future_work}. Surely, for very small $\xi$ the 1D scenario will break down.  Indeed, a planar crystalline droplet has been predicted recently  for $\xi=0$\cite{Langen2}. In this way, the ellipticity ($\xi$) serves as an efficient parameter to control the effective dimension of polar molecules. In future, it will be interesting to study $\xi$-driven dimensional crossover of distinct crystalline states, as well as the effect of dual microwave fields in bosonic molecules\cite{Shi4, review_dualmicrowavewave}.  

Data that support the findings of this article are openly available\cite{data}.

{\it Acknowledgements.} We thank Mingyuan Sun, Ruijin Liu, Peng Zhang and Tao Shi for useful discussions, and Andreas Schindewolf, Bo Zhao and Pengjun Wang for helpful feedback on the manuscript. This work is supported by National Natural Science Foundation of China (92476104, 12525412, 12134015) and Quantum Science and Technology-National Science and Technology Major Project (2024ZD0300600). T.S. thanks support from the Postdoctoral Fellowship Program of CPSF (Grants No. GZC20232945).

\clearpage

\onecolumngrid
\vspace*{1cm}
\begin{center}
{\large\bfseries Supplementary Materials}
\end{center}
\setcounter{figure}{0}
\setcounter{equation}{0}
\renewcommand{\figurename}{Fig.}
\renewcommand{\thefigure}{S\arabic{figure}}
\renewcommand{\theequation}{S\arabic{equation}}

In this supplementary material, we provide more details on the derivations of effective 1D models and the   properties of tetratomic and hexatomic systems. 

\section*{I.\ \ \ Derivations of effective 1D models}

\subsection{A. Two molecules}

Expanding $V({\cp r})$ up to the lowest fluctuation order $\sim \delta\theta^2, \delta\phi^2$, we have
\begin{equation}
V({\cp r})=-\frac{4C_3}{r^3}+\left(\frac{6C_3}{r^3} + \frac{4C_6}{r^6}\right) (\delta\theta^2+\delta\phi^2). \label{V2}
\end{equation}
Together with the kinetic term,  $H_{\rm kin}=-\frac{\hbar^2}{mr^2}\frac{\partial}{\partial r}\left(r^2\frac{\partial}{\partial r}\right)-\frac{\hbar^2}{mr^2}\left(\frac{\partial^2}{\partial \delta\theta^2} + \frac{\partial^2}{\partial \delta\phi^2}\right)$, 
 $H^{(2)}$ is then reduced to 
\begin{equation}
H^{(2)}({\cp r})=-\frac{\hbar^2}{mr^2}\frac{\partial}{\partial r}\left(r^2\frac{\partial}{\partial r}\right)-\frac{4C_3}{r^3}+\left[-\frac{\hbar^2}{mr^2}\frac{\partial^2}{\partial \delta\theta^2} +\left(\frac{6C_3}{r^3} + \frac{4C_6}{r^6}\right) \delta\theta^2 \right]+ \left[-\frac{\hbar^2}{mr^2}\frac{\partial^2}{\partial \delta\phi^2}+ \left(\frac{6C_3}{r^3} + \frac{4C_6}{r^6}\right) \delta\phi^2\right]. \label{H2}
\end{equation}
Apparently, the angular part of $H^{(2)}({\cp r})$ is composed by two independent harmonic oscillators with respect to $\delta\theta$ and $\delta\phi$, which give rise to discrete energy levels $\epsilon_{\nu}=(\nu+1)\sqrt{\frac{4\hbar^2}{m}\left(\frac{6C_3}{r^5} + \frac{4C_6}{r^8}\right)}$. Note that in expanding $H_{\rm kin}$, we have neglected the term $\sim \delta\theta\frac{\partial}{\partial\delta\theta}$ whose expectation value ($\sim$constant under $\nu=0$) is much smaller than $\langle\frac{\partial^2}{\partial\delta\theta^2}\rangle \sim 1/r^2$ at short $r$, and thus it belongs to high-order fluctuations. 

Further writing $\Psi^{(2)}$ under adiabatic representation, and neglecting off-diagonal couplings between different $\nu$, we get the reduced 1D equation for the ground state ($\nu=0$) as Eq.(3) in the main text.

\subsection{B. A light molecule interacting with two heavy ones: Born-Oppenheimer limit}

Here we derive the effective 1D model for the light molecule (${\cp r}$) moving around two heavy ones with relative distance ${\cp R}=R\hat{y}$. The interaction potential felt by the light molecule is composed by $V({\cp r}_{+})$ and $V({\cp r}_{-})$, where ${\cp r}_{\pm}={\cp r}\pm{\cp R}/2$ is the relative coordinate between a heavy-light pair.  Following the expansion in Eq.(\ref{V2}), we have 
\begin{equation}
V({\cp r}_{+})+V({\cp r}_{-})= \sum_{\sigma=\pm} \left( -\frac{4C_3}{r_{\sigma}^3}+\left(\frac{6C_3}{r_{\sigma}^3} + \frac{4C_6}{r_{\sigma}^6}\right) (\delta\theta_{\sigma}^2+\delta\phi_{\sigma}^2)  \right), \label{V_BO}
\end{equation}
where $r_{\pm}=|{\cp r}_{\pm}|$, and $\delta\theta_{\pm},\delta\phi_{\pm}$ represent the angular fluctuations of ${\cp r}_{\pm}$ from $y$ direction. For ${\cp R}$ along $y$, we have $\delta\theta_{\pm}=r\delta\theta/r_{\pm}$ and $\delta\phi_{\pm}=r\sin\phi_0\delta\phi/r_{\pm}$. Now we can simplify (\ref{V_BO}) as
\begin{equation}
V({\cp r}_{+})+V({\cp r}_{-})=-\frac{4C_3}{|y_{+}|^3} -\frac{4C_3}{|y_{-}|^3} + (\delta\theta^2 + \delta\phi^2)Ar^2, \label{V_final}
\end{equation}
with $y_{\pm}=y\pm R/2$ ($y=r\sin\phi_0$) and
\begin{equation}
A=6C_3\left(\frac{1}{|y_-|^5}+\frac{1}{|y_+|^5}\right) +4C_6\left(\frac{1}{|y_-|^8}+\frac{1}{|y_+|^8}\right). \label{A}
\end{equation}
Note that in obtaining (\ref{V_final}), we have neglected the angular fluctuations of radial distance $r_{\pm}$ and replaced it directly with $|y_{\pm}|$. 
By doing this, one can ensure that (\ref{V_final}) correctly reproduce the effective potential of a heavy-light pair at $r_{\pm}\rightarrow 0$, i.e.,  whenever the light molecule approaches a heavy one. 

Together with the kinetic term, the light Hamilonian $H_L({\cp r})$ can be expanded as
\begin{eqnarray}
H_L({\cp r})&=&-\frac{\hbar^2}{2mr^2}\frac{\partial}{\partial r}\left(r^2\frac{\partial}{\partial r}\right)-4C_3\left(\frac{1}{|y_-|^3}+\frac{1}{|y_+|^3} \right)-\frac{\hbar^2}{2m r^2} \left(
	\frac{\partial^2}{\partial \delta\theta^2} + \frac{\partial^2}{\partial \delta\phi^2}
	\right)+(\delta\theta^2 + \delta\phi^2)Ar^2, \label{H2}
\end{eqnarray}
Again, the angular part of (\ref{H2}) is composed by two independent harmonic oscillators, with discrete energy level $\epsilon_{\nu}=(\nu+1)\sqrt{2\hbar^2A/m}$. 

Writing the wavefunction of light molecule under adiabatic representation
\begin{equation}
\Psi_L({\cp r})=\frac{1}{r}\sum_{\nu} F_{\nu}(y) \psi_{\nu}(y;\delta\theta,\delta\phi), \label{psiL}
\end{equation}
and neglecting off-diagonal couplings between different $\nu$, we obtain the ground state equation for the light molecule effectively moving along $y$: 
\begin{equation}
\left( - \frac{\hbar^2}{2m} \frac{\partial^2}{\partial y^2} + U_L(y) \right) F_{0}(y) = V_{\rm BO}F_{0}(y), \label{Heff_L}
\end{equation}
where $U_L$ follows Eq.(6) in the main text with $u_L(y)=\sqrt{2\hbar^2A/m}$.

\subsection{C. Three identical molecules}

The relative coordinates $\{ {\cp r},\bm{\rho}\}$ of three identical molecules are assumed with small angular fluctuations from $y$ direction: $\{\delta\theta_{r},\delta\phi_{r}, \delta\theta_{\rho},\delta\phi_{\rho}\}$. The kinetic term of $H^{(3)}$ can be expanded as
\begin{equation}
-\frac{\hbar^2}{m} (\nabla^2_{\cp r}+\nabla^2_{\bm{\rho}})\rightarrow -\frac{\hbar^2}{mr^2}\frac{\partial}{\partial r}\left(r^2\frac{\partial}{\partial r}\right) -\frac{\hbar^2}{m\rho^2}\frac{\partial}{\partial \rho}\left(\rho^2\frac{\partial}{\partial \rho}\right) -\frac{\hbar^2}{m r^2} \left(
	\frac{\partial^2}{\partial \delta\theta_r^2} + \frac{\partial^2}{\partial \delta\phi_r^2}\right) -\frac{\hbar^2}{m \rho^2} \left(
	\frac{\partial^2}{\partial \delta\theta_{\rho}^2} + \frac{\partial^2}{\partial \delta\phi_{\rho}^2}\right).
\end{equation}
Following the strategy in above section, the interaction part of $H^{(3)}$ can be expanded as (here ${\cp r}_{\pm}=\frac{\cp r}{2}\pm\frac{\sqrt{3}{\bm{\rho}}}{2}$):
\begin{equation}
V({\cp r})+V({\cp r}_+)+V({\cp r}_-) \rightarrow -4C_3\Big(
	\frac{1}{|y_r|^3} + \frac{1}{|y_{+}|^3} + \frac{1}{|y_{-}|^3}
	\Big) + \sum_{\sigma=r,+,-} \left(\frac{6C_3}{|y_{\sigma}|^3} + \frac{4C_6}{|y_{\sigma}|^6}\right) (\delta\theta_{\sigma}^2+\delta\phi_{\sigma}^2),
		\label{V3}
\end{equation}
where $y_{\pm}=\frac{y_r}{2}\pm\frac{\sqrt{3}{y_{\rho}}}{2}$, with  $y_r=r\sin\phi_{0,r}$, $y_{\rho}=\rho\sin\phi_{0,\rho}$ the projection of ${\cp r},\ \bm{\rho}$ along $y$ direction; $\delta\theta_{\pm}=(\frac{|y_r|}{2}\delta\theta_r \pm \frac{\sqrt{3}|y_{\rho}|}{2}\delta\theta_{\rho})/|y_{\pm}|$ and $\delta\phi_{\pm}=(\frac{y_r}{2}\delta\phi_r \pm \frac{\sqrt{3}y_{\rho}}{2}\delta\phi_{\rho})/|y_{\pm}|$ are the angular fluctuations of ${\cp r}_{\pm}$. Finally we have
\begin{equation}
V({\cp r})+V({\cp r}_+)+V({\cp r}_-) \rightarrow -4C_3\Big(
	\frac{1}{|y_r|^3} + \frac{1}{|y_{+}|^3} + \frac{1}{|y_{-}|^3}
	\Big) + c (\delta\theta_r^2 + \delta\phi_r^2) 
	+  d (\delta\theta_{\rho}^2 + \delta\phi_{\rho}^2) 
	+  e_{\theta} \delta\theta_r \delta\theta_{\rho}
	+e_{\phi} \delta\phi_r \delta\phi_{\rho}  \label{V4}
\end{equation}
with 
\begin{eqnarray}
	c &=& C_3\,\Big[
	\frac{6}{|y_r|^3} + \frac{3y_r^2}{2} \Big(
	\frac{1}{|y_{+}|^5} + \frac{1}{|y_{-}|^5}
	\Big)	\Big] + C_6 \, \Big[
	\frac{4}{|y_r|^6} + y_r^2\Big(
	\frac{1}{|y_{+}|^8} + \frac{1}{|y_{-}|^8}
	\Big)
	\Big]\notag\\
	d &=& C_3\,\Big[
	\frac{9 y_\rho^2}{2} \Big(
	\frac{1}{|y_{+}|^5} + \frac{1}{|y_{-}|^5}
	\Big)	\Big]
	+ C_6 \, 3y_\rho^2
	\Big(
	\frac{1}{|y_{+}|^8} + \frac{1}{|y_{-}|^8}
	\Big)\notag\\	
	e_{\theta}  &=& C_3 \, 3\sqrt{3} |y_ry_\rho| \Big(
	\frac{1}{|y_{+}|^5} - \frac{1}{|y_{-}|^5}
	\Big)
	+ C_6\, 2\sqrt{3} |y_ry_\rho|\Big(
	\frac{1}{|y_{+}|^8} - \frac{1}{|y_{-}|^8}
	\Big)\notag\\	
	e_{\phi} &=& C_3 \, 3\sqrt{3} y_r y_\rho \Big(
	\frac{1}{|y_{+}|^5} - \frac{1}{|y_{-}|^5}
	\Big)
	+ C_6\, 2\sqrt{3} y_r y_\rho\Big(
	\frac{1}{|y_{+}|^8} - \frac{1}{|y_{-}|^8}
	\Big).
\end{eqnarray}
Similar as the heavy-heavy-light case, in obtaining (\ref{V3}) we have replaced the radial distance $|{\cp r}_{\pm}|$ directly with $|y_{\pm}|$. This not only ensures the correct asymptotic  potential whenever pairwise molecules come closer,  but also guarantees the full exchange symmetry of reduced potential under $y_i\leftrightarrow y_j$ (or mutual exchange between $\{y_r,y_+,y_-\}$).

From (\ref{V4}), we can see that the expansion involves off-diagonal couplings as $\delta\theta_r \delta\theta_{\rho}$ and $\delta\phi_r\delta\phi_{\rho}$, which originate from the angular fluctuations of ${\cp r}_{\pm}$. So we have two independent groups of fluctuations:  $\{\delta\theta_r, \delta\theta_{\rho}\}$ and $\{\delta\phi_r, \delta\phi_{\rho}\}$. For each group, we need to diagonalize it to obtain zero-point energy, which contributes to the effective 1D force.  The diagonalization can be done as follows. In general, any bilinear Hamiltonian
\begin{equation}
H(x,y)=-a \frac{\partial^2}{\partial x^2} -b\frac{\partial^2}{\partial y^2} + c\,x^2 + d\,y^2 + e\,xy \label{e1}
\end{equation}
can be diagonalized into the form
\begin{equation}
H(x,y)=-\tilde{a} \frac{\partial^2}{\partial \tilde{x}^2} -\tilde{b} \frac{\partial^2}{\partial \tilde{y}^2} + \tilde{c}\,\tilde{x}^2 + \tilde{d}\,\tilde{y}^2, \label{e2}
\end{equation}
where $\tilde{x},\tilde{y}$ are linear combinations of $x,y$. After straightforward algebra, we obtain  the  zero-point energy
\begin{equation}
\epsilon_0=\sqrt{\tilde{a}\tilde{c}}+\sqrt{\tilde{b}\tilde{d}}=\sum_{\alpha=\pm 1} \sqrt{\frac{ac+bd}{2}+\frac{\alpha}{2}\sqrt{(ac-bd)^2+abe^2}}. \label{e3}
\end{equation}

For three identical molecules with a given projection order along $y$, such as $y_1<y_2<y_3$, we can write down their wavefunction as
\begin{equation}
\Psi^{(3)}({\cp r},\bm{\rho})=\frac{1}{r\rho}\sum_{\nu} F_{\nu}(y_r,y_{\rho}) \psi_{\nu}(y_r, y_{\rho}; \delta\theta_r,\delta\phi_r,\delta\theta_{\rho},\delta\phi_{\rho}). \label{psi3}
\end{equation}
Neglecting all off-diagonal couplings between different $\nu$, we obtain the reduced 1D equation for ground state ($\nu=0$):
\begin{equation}
\left( - \frac{\hbar^2}{m} \frac{\partial^2}{\partial y_r^2} - \frac{\hbar^2}{m} \frac{\partial^2}{\partial y_{\rho}^2}  + U^{(3)}(y_r,y_{\rho}) \right) F_{0}(y_r,y_{\rho}) = E_{\rm 1D}^{(3)}F_{0}(y_r,y_{\rho}),
\end{equation}
where $U^{(3)}$ follows Eq.(7) in the main text, with $u^{(3)}$ the zero-point energy by diagonalizing the fluctuations (following strategy in Eqs.(\ref{e1},\ref{e2},\ref{e3})).   

\section*{II.\ \ \ Tetratomic system: Comparison between exact 3D and effective 1D results }

We have exactly solved the tetratomic bound states of two molecules by expanding their relative wavefuntion as
\begin{equation}
\Psi^{(2)}({\bf r}) = \int_0^{+\infty}d k \sum_{lm} C_{lm}(k) \phi_k^l(r) Y_{lm}(\theta,\phi),\label{klm}
\end{equation}
where the radial basis is $\phi_k^l(r)\equiv \sqrt{\frac{2}{\pi}} k j_l(kr)$ with $j_l$ the spherical Bessel function of the first kind. Eigen-energy $E^{(2)}$ and $\{C_{lm}(k)\}$ can be obtained by diagonalizing  $H^{(2)}({\cp r})=-\hbar^2\nabla_{\cp r}^2/m +V({\cp r})$ in $\{klm\}$ space. Solutions of bosonic (fermionic) systems are associated with even (odd)  $l$ due to symmetry requirement.

In Fig.~2 and Fig.~3(d) of the main text, we have compared the exact results with those from the effective 1D model, and we can see that the effective 1D model well reproduces the binding energies and density correlation functions of tetratomic bound states. Below we provide more details on the comparison between exact 3D and effective 1D results. 

\begin{figure}[h]
	\includegraphics[width=0.5\linewidth]{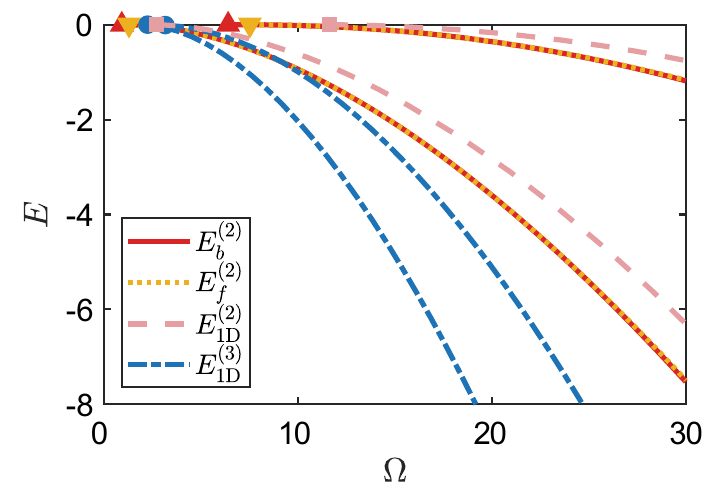}
	\caption{Magnified plot of Fig.2 in the main text at small binding energies. The  critical microwave field for the emergence of various bound states is marked with according color. 
		The units of $E$ and $\Omega$ are $E_u$ and $E_u/\hbar$.} 
	\label{fig_magfig1}
\end{figure}

In Fig.\ref{fig_magfig1}, we show a magnified plot of Fig.~2 in the main text, focusing on shallow bound states at small $\Omega$. The location of critical $\Omega_c$ for their individual emergence is marked accordingly. For tetratomic bound states, the exact binding energies of both bosonic (red solid) and fermionic  (yellow dot) are generally deeper than those predicted by the 1D model (pink dashed, identical to bosonic and fermionic systems); accordingly, the exact solution requires a smaller $\Omega_c$  as compared to that from 1D model. The discrepancy between 1D and exact results  is largely due to the omission of higher-order angular fluctuations in the effective model. Moreover, we can see from Fig.\ref{fig_magfig1} that the bosonic and fermionic systems have almost identical energies across a wide range of $\Omega$, confirming the Bose-Fermi duality as inferred from effective 1D model.

An even transparent comparison between 1D and 3D tetratomic states is from their wavefunctions. For a direct comparison, we define the reduced 3D wavefunction along $y$  as
\begin{equation}
\tilde{\Psi}^{(2)}(y)\equiv \sqrt{\int dx dz |\Psi^{(2)}({\cp r})|^2} \times {\rm Sgn}(\Psi^{(2)}({\cp r})). \label{psi_1d}
\end{equation}
Here ${\rm Sgn}(x)$ is a sign function, which is $1$ ($-1$) when $x$ is positive (negative); $\Psi^{(2)}({\cp r})$ is (normalized) 3D wavefunction of tetratomic state, and $\tilde{\Psi}^{(2)}(y)$ is its reduced version along $y$.

\begin{figure}[h]
	\includegraphics[width=0.45\linewidth]{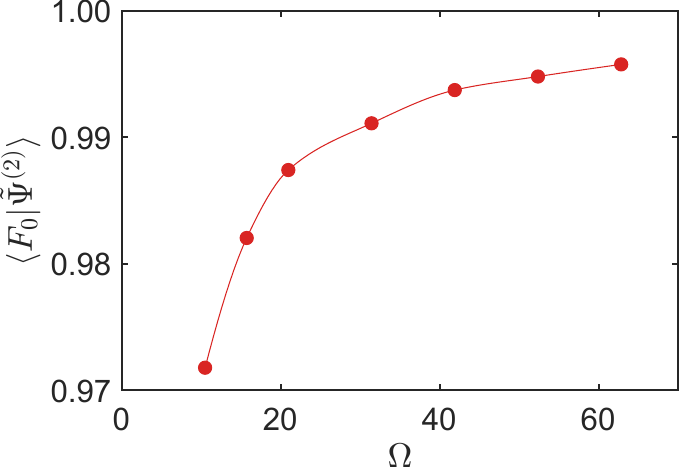}
	\caption{Wavefunction overlap between the lowest tetratomic states from effective 1D treatment ($F_0(y)$) and from exact 3D solutions $(\tilde{\Psi}^{(2)}(y)$, see (\ref{psi_1d})). The unit of $\Omega$ is $E_u/\hbar$.} 
	\label{figs2}
\end{figure}

In Fig.~\ref{figs2}, we plot out the overlap $\langle{F_0}|{\tilde{\Psi}^{(2)}}\rangle$ as a function of $\Omega$, with $F_0(y)$ the wavefunction from effective 1D model. One can see an excellent agreement between the two wavefunctions for a wide range of $\Omega$, with overlap exceeding $97\%$ for $\hbar\Omega/E_u>10$. This is why the correlation functions, shown by $G_2(y)$ in Fig.3(d) of the main text, fit so well with each other.   


\begin{figure}[h]
	\includegraphics[width=0.9\linewidth]{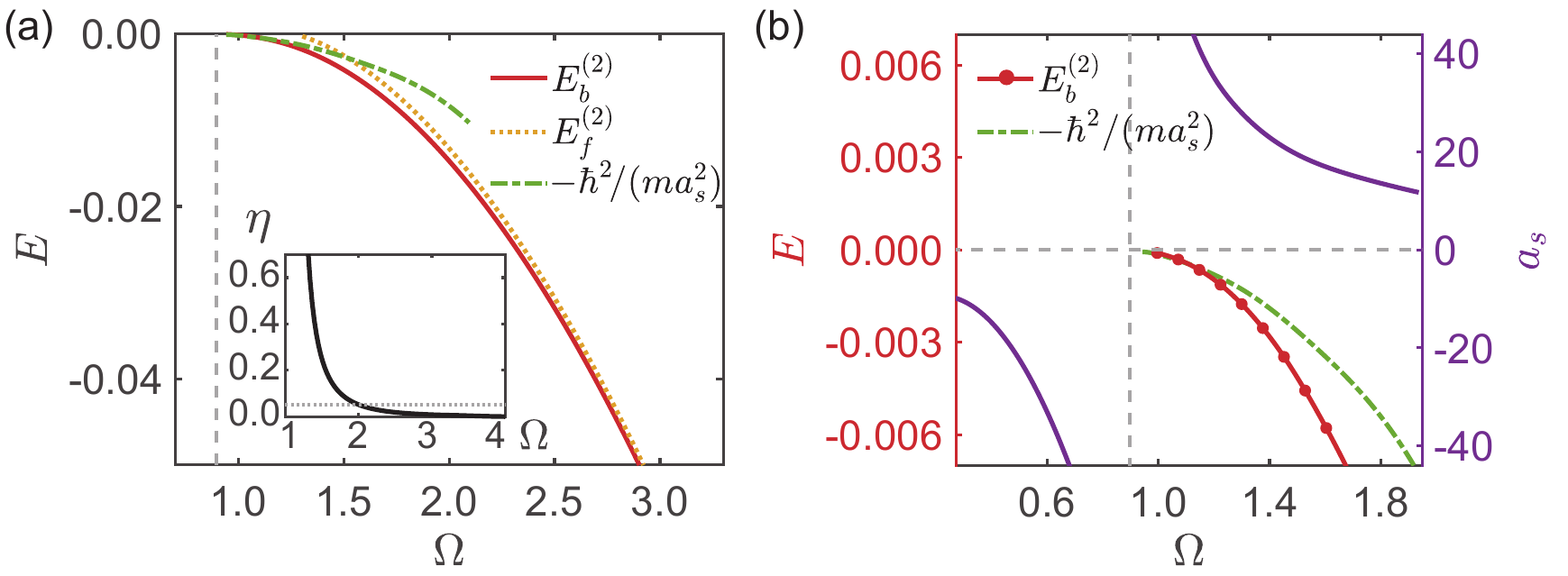}
	\caption{(a) Exact tetratomic binding energies of bosonic ($E_b^{(2)}$, red solid) and fermionic ($E_f^{(2)}$, yellow dot) molecules as functions of $\Omega$ near their emergence. The exact $E_b^{(2)}$ is compared with s-wave prediction $-\hbar^2/(ma_s^2)$ (green dashed-dot), with $a_s$ given in (b).  Inset shows the relative difference  $\eta=|E_b^{(2)}-E_f^{(2)}|/|E_b^{(2)}+E_f^{(2)}|$, and the  horizontal line locates at  $\eta=5\%$. (b) Exact $E_b^{(2)}$ (red line with circles), s-wave prediction  $-\hbar^2/(ma_s^2)$ (green dashed-dot), and s-wave scattering length $a_s$ (purple solid) in a smaller region of $\Omega$ near resonance. Vertical lines in (a) and (b) locate the resonance position where $a_s\rightarrow\infty$. 
	The units of length, energy and $\Omega$ are $l_u$, $E_u$ and $E_u/\hbar$ respectively.} 
	\label{fig_res}
\end{figure}

Now we look into the very shallow tetratomic bound states near their emergence.  In Fig.\ref{fig_res}(a), we show even magnified plots of $E_b^{(2)}$ and $E_f^{(2)}$ close to their emergence. We can see that the bosonic bound state ($E_b^{(2)}$) emerges at a smaller $\Omega_c$ than the fermionic one ($E_f^{(2)}$). However, as increasing $\Omega$, both $E_b^{(2)}$ and $E_f^{(2)}$ get deeper and they soon become very close to each other.  As shown in the inset of Fig.\ref{fig_res}(a), the relative difference between $E_b^{(2)}$ and $E_f^{(2)}$ quickly reduces to below $5\%$ for $\hbar\Omega/E_u>2$, signifying the entrance of effective 1D regime.

In fact, the emergence of $E_b^{(2)}$ is associated with an s-wave resonance as previously studied in dipolar systems\cite{You, Blume2, Cavagnero, Qi_Zhai}, where the s-wave scattering length $a_s\rightarrow\infty$. Nearby the resonance, $E_b^{(2)}\sim -\hbar^2/(ma_s^2)$. Here, we have extracted $a_s$ from  the phase shift of asymptotic s-wave wavefunction at long range, i.e., $a_s=-\lim_{k\rightarrow 0} \tan \delta_k/k$.  In Fig.\ref{fig_res}(b), we plot out $E_b^{(2)}$ (red solid), its s-wave prediction $-\hbar^2/(ma_s^2)$ (green dashed-dot), and s-wave scattering length $a_s$ (purple solid)  near resonance. We can see that $E_b^{(2)}$ emerges exactly at $a_s\rightarrow\infty$ and near resonance it well follows the s-wave prediction. This confirms the nature of very shallow bound states in bosonic molecules as the s-wave bound states in 3D, rather than the states predicted by effective 1D model. Similarly, the shallow  bound states in fermionic molecules should be p-wave bound states in 3D, but not follow effective 1D description either. Therefore, the bosonic and fermionic bound states are qualitatively different nearby their emergence, where the effective 1D model fails to describe them.

Finally we briefly comment on possible Efimov physics near s-wave resonance. According to Ref.\cite{Greene1}, the Efimov physics in dipolar systems emerge at length scale $l_d\ll r\ll a_s$. 
Given the current molecular systems  with very large $l_d\gtrsim 10^4 a_0$, Efimov trimers are associated with even larger length scale and thus expected to be very shallow and only exist in a narrow window near resonance ($a_s\rightarrow \infty$). As shown in Fig.\ref{fig_res}(b), the condition $|a_s|> l_d(=20l_u)$ is satisfied only very close to resonance with $\hbar\Omega\in (0.5, 1.5)E_u$, where the tetratomic binding energy is exceedingly shallow $|E_b^{(2)}|\le 0.004E_u$. Therefore, we expect Efimov trimers to be equally shallow within this narrow regime, which are  extremely hard to explore in realistic experiment. This also explains why we do not observe Efimov physics in the Born-Oppenheimer limit (Fig.4 in the main text), i.e., the scattering length therein does not satisfy the stringent condition $a_s\gg l_d$. 


\section*{III.\ \ \ Hexatomic system}

\subsection{A. Born-Oppenheimer potential for heavy-heavy-light system 
}

For the light molecule interacting with two heavy molecules localized at $\pm{\cp R}/2$, we  have exactly solved $H_L({\cp r})\Psi_L({\cp r})=V_{\rm BO}({\cp R})\Psi_L({\cp r})$ by expanding the light wavefunction $\Psi_L({\cp r})$ in $\{klm\}$ space (similar to Eq.\ref{klm}). The eigen-energy $V_{\rm BO}({\cp R})$ can be seen as the effective potential between heavy objects induced by the movement of the light object.

First, we will focus on the case when  ${\cp R}$ is along $y$ direction i.e., ${\bf R}=R\hat{\bf y}$. In this case we obtain two eigen-levels of $V_{\rm BO}$, as shown by the red and blue triangles in  Fig.4 of the main text. These two levels correspond to two orthogonal states of light molecule, i.e., when it lies in-between (blue triangles) or outside (red triangles) of two heavy molecules. As $R$ decreases, the two $V_{\rm BO}$ undergo a first-order level crossing,  and the ground state of light molecule switches from the in-between configuration to the outside one.

\begin{figure}[hbpt]     
	\centering
	\includegraphics[width=0.98\linewidth]{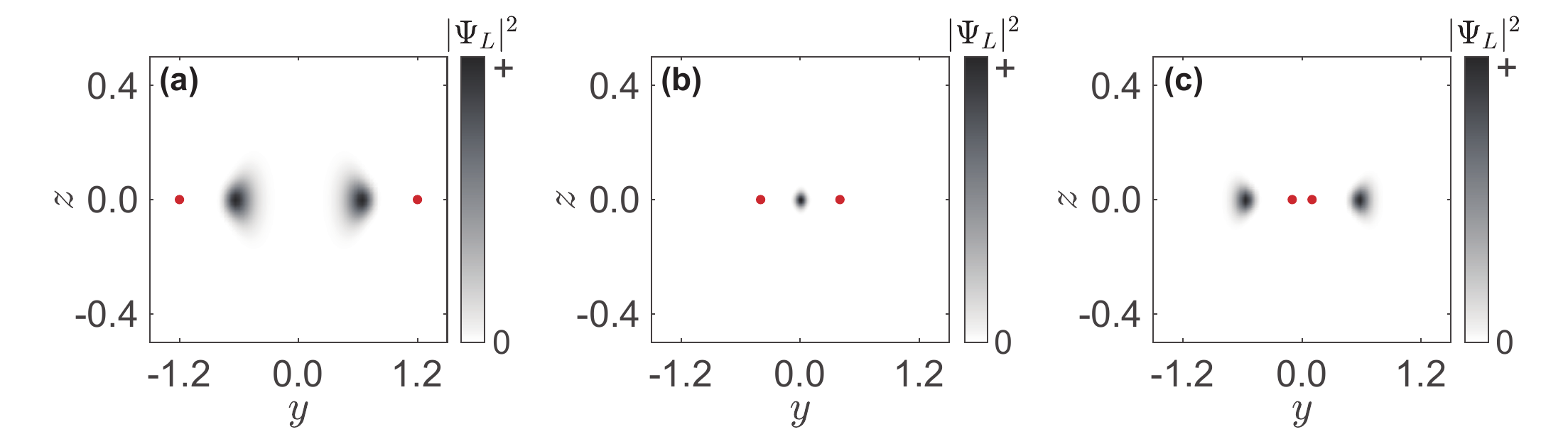}
	\caption{Probability of the light molecule $|\Psi_L({\bf r})|^2$ (at $x=0$)  when two heavy ones stays along $y$ direction with distance $R$ (located by two red points). (a) and (b) are for the light molecule staying in-between two heavy ones, with $R=2.4$ and $0.8$ respectively. (c) is for the light molecule staying outside of two heavy ones, with $R=0.2$. Here $\hbar\Omega/E_u = 52$, and the length unit is $l_u$.}
	\label{figs4} 
\end{figure}

For the in-between configuration, $V_{\rm BO}$ is attractive at large $R$ and repulsive at short $R$, similar to the behavior of $U^{(2)}$ in Eq.~(4) of the main text. 
At large $R$, the light molecule tend to stay near the heavy ones and form light-heavy dimers, see  density profile in Fig.~\ref{figs4}(a). At short $R$, the light molecule stays around ${\cp r}\sim0$ (see  Fig.~\ref{figs4}(b)) and thus feels a strong heavy-light repulsion due to the induced $|r\pm R/2|^{-4}$ repulsion (as illustrated in Fig.1 of the main text). However, for the outside configuration, at short $R$ the light molecule can avoid this short-range repulsion by staying away from two heavy ones, see Fig.~\ref{figs4}(c). In this case, $V_{\rm BO}$ monotonically decreases as $R$ gets smaller, reflecting the enhanced attraction between heavy-heavy molecules due to the motion of light one. At $R\rightarrow 0$, $V_{\rm BO}$ saturates at a negative value corresponding to the eigen-energy of light molecule under twice of $V({\cp r})$.  

\begin{figure}[hbpt]     
	\centering
	\includegraphics[width=0.98\linewidth]{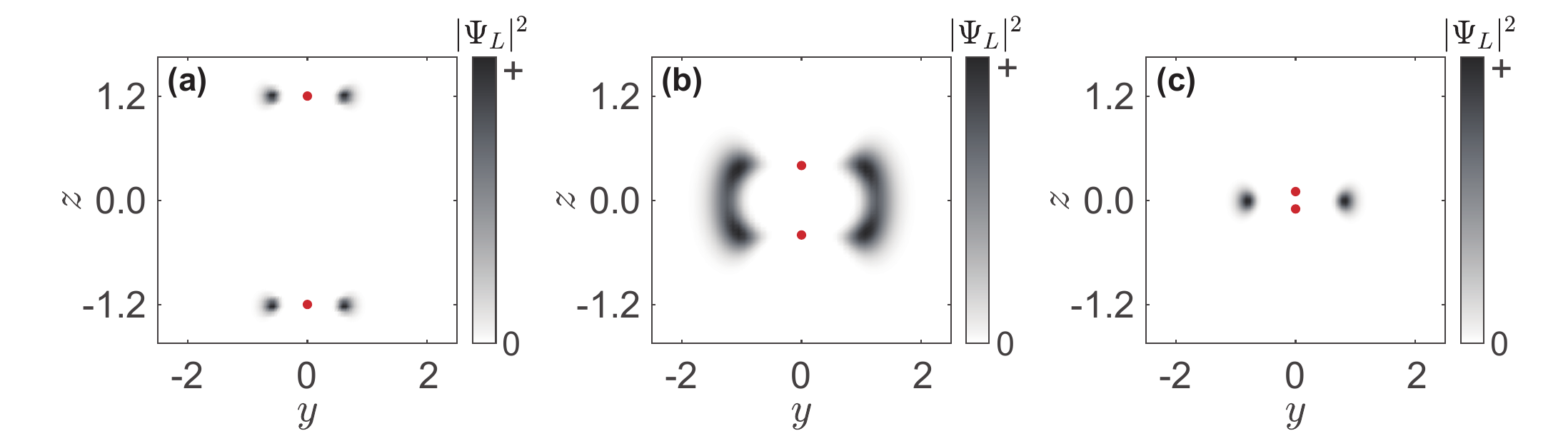}
	\caption{Probability of the light molecule $|\Psi_L({\bf r})|^2$ (at $x=0$) when the two heavy ones stays along $z$ direction with distance $R=2.4$ (a), $0.8$ (b) and $0.2$ (c). Red points denote the locations of two heavy molecules. Here $\hbar\Omega/E_u = 52$, and the length unit is $l_u$.}
	\label{figs5} 
\end{figure}

Compared to above case, $V_{\rm BO}$ behaves very differently if ${\cp R}$ orientates along $x$ or $z$. Note that we have $V_{\rm BO}(R\hat{x})=V_{\rm BO}(R\hat{z})$ due to the symmetry of bare interaction potential $V({\cp r})$ under a $\pi/4$ rotation around $y$-axis ($\hat{x}\rightarrow\hat{z}$, $\hat{z}\rightarrow-\hat{x}$). Therefore in the following we just take ${\cp R}=R\hat{z}$ for example. The typical distributions of light molecule in this case are shown in Fig.\ref{figs5}(a,b,c) for several different $R$. For large $R$,  
the light molecule stays near the two heavy ones to form light-heavy dimers, and the two dimers are both orientated along $y$ direction, see Fig.\ref{figs5}(a). As $R$ is reduced, the density peaks of light molecule get closer along $z$ direction (Fig.\ref{figs5}(b)) and finally at very small $R$ they merge into two peaks at $z\sim 0$ and distributed along $y$ (Fig.\ref{figs5}(c)). During this process, $V_{\rm BO}$ decreases continuously and there is no sharp transition as in the case of ${\cp R}=R\hat{y}$.  

The behavior of $V_{\rm BO}$ at large $R$ can be well predicted  by the mean-field energy between this heavy-light dimer and the rest heavy molecule. Specifically, we take four dimer bases $\{\Psi^{(2)}_{-}({\bf r+\frac{R}{2}}), \Psi^{(2)}_{+}({\bf r+\frac{R}{2}}), \Psi^{(2)}_{-}({\bf r-\frac{R}{2}}), \Psi^{(2)}_{+}({\bf r-\frac{R}{2}})\}$, where ${\bf r\pm\frac{R}{2}}$ is the relative distance between the light molecule and two heavy ones, and the subscript $\pm$ refers to the light molecule one staying at $+y$ or $-y$ side of the heavy one. Then the Hamiltonian $H_L({\bf r})$  can be expanded under above bases as a $4\times 4$ matrix. For instance, the typical diagonal elements are:
\begin{eqnarray}
(H_L)_{11} &=&\int d{\bf r}\ {\Psi^{(2)}}^*_-({\bf r+\frac{R}{2}})H_L\Psi^{(2)}_-({\bf r+\frac{R}{2}}) = E^{(2)} + \int d{\bf r} \ |\Psi^{(2)}_-({\bf r})|^2 V({\bf r-R});\\
(H_L)_{22} &=&\int d{\bf r}\ {\Psi^{(2)}}^*_+({\bf r+\frac{R}{2}})H_L\Psi^{(2)}_+({\bf r+\frac{R}{2}}) = E^{(2)} + \int d{\bf r} \ |\Psi^{(2)}_+({\bf r})|^2 V({\bf r-R}),
\end{eqnarray}
and the typical off-diagonal elements are
\begin{eqnarray}
(H_L)_{12} &=&\int d{\bf r}\ {\Psi^{(2)}}^*_-({\bf r+\frac{R}{2}})H_L\Psi^{(2)}_+({\bf r+\frac{R}{2}}) = \int d{\bf r}\ {\Psi^{(2)}}^*_-({\bf r})(E^{(2)}+V({\bf r-R}))\Psi^{(2)}_+({\bf r});\\
(H_L)_{13} &=&\int d{\bf r}\ {\Psi^{(2)}}^*_-({\bf r+\frac{R}{2}})H_L\Psi^{(2)}_-({\bf r-\frac{R}{2}}) = \int d{\bf r}\ {\Psi^{(2)}}^*_-({\bf r}+{\bf R})(E^{(2)}+V({\bf r+R}))\Psi^{(2)}_-({\bf r}).
\end{eqnarray}
Because of the vanishing overlap between different bases functions, all off-diagonal elements are extremely small at large $R$ and can be neglected. Therefore only the diagonal terms contribute to $V_{BO}$, which correspond to the dimer energy $E^{(2)}$ shifted by a mean-field interaction energy between the heavy-light dimer and the rest heavy molecule. Such mean-field shifts are shown as dotted lines in Fig.4 of the main text, which well fit the exact numerical results at large $R$. 

\subsection{B. Hexatomic bound states of three identical molecules}

\begin{figure}[h]
	\includegraphics[width=0.4\linewidth]{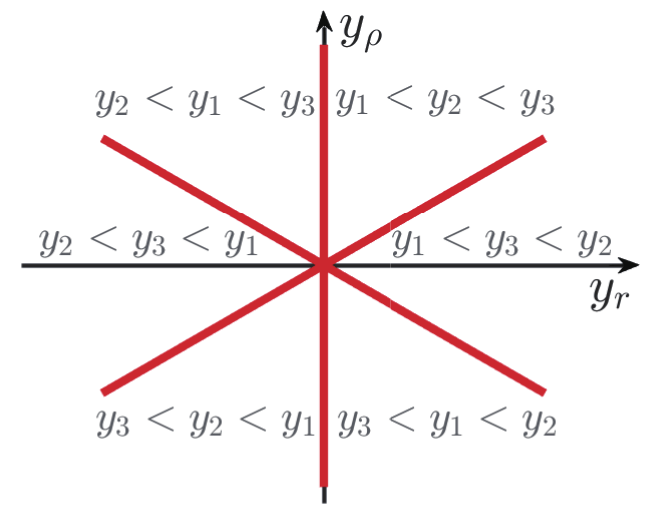}
	\caption{Symmetry axes (red lines) and  specific orders of three identical molecules along $y$ direction. Here $\{y_1,y_2,y_3\}$ are the coordinates of three molecules, and $\{y_r, y_{\rho}\}$ are the relative coordinates defined in (\ref{relative_y}).}  \label{fig_symmetry}
\end{figure}

In this subsection we discuss the hexatomic bound states of three identical molecules under the effective 1D model. For three molecules at coordinates $\{y_1,y_2,y_3\}$, in the center-of-mass frame they can be described by two relative coordinates 
\begin{equation}
y_r\equiv y_2-y_1,\ \ \ y_{\rho}\equiv \frac{2}{\sqrt{3}}\left(y_3-\frac{y_1+y_2}{2}\right). \label{relative_y}
\end{equation} 
Due to the orthogonality of these coordinates, the kinetic term can be fully decoupled to $-\frac{\hbar^2}{m}(\frac{\partial^2}{\partial y_r^2}+\frac{\partial^2}{\partial y_{\rho}^2})$. The order of $\{y_1,y_2,y_3\}$ on the 1D line can also be identified by separated regions in $(y_r,y_{\rho})$ plane. As shown in Fig.\ref{fig_symmetry},  $(y_r,y_{\rho})$ plane can be equally divided to six pieces and each piece corresponds to a given order of three particles along $y$ direction. According to the (bosonic or fermionic) statistics, the wavefunction $\Psi^{(3)}(y_r,y_{\rho})$   should respect (symmetric or antisymmetric) reflection symmetry around any of three axes:  $y_r=0$ and $y_{\rho}=\pm y_r/\sqrt{3}$ as highlighted by red lines in Fig.\ref{fig_symmetry}. Such reflection symmetry is equivalent to the exchange symmetry of $\Psi^{(3)}$ under $y_i\leftrightarrow y_j$. 
To ensure this symmetry, the interaction potential $U^{(3)}(y_r,y_{\rho})$ should be fully symmetric around these axes, which we have confirmed numerically under the effective 1D framework. 

Alternatively, we can define another set of relative coordinates $\{ y_{12}\equiv y_2-y_1,  y_{23}\equiv y_3-y_2\}$ to replace $\{y_r,y_{\rho}\}$. The advantage of using   $\{ y_{12}, y_{23}\}$ is that it facilitates  the construction of hexatomic bound states from tetratomic ones. To be concrete, let us consider three molecules aligning along $y$ with a particular order $y_1<y_2<y_3$, which means $y_{12}>0$ and  $y_{23}>0$. Together with the center-of-mass coordinate $R\equiv (y_1+y_2+y_3)/3$, one can build up the relation between $\{y_1,y_2,y_3\}$ and $\{y_{12},y_{23},R\}$, and transform the Hamiltonian $\tilde{H}_{\rm 1D}^{(3)}$ as 
\begin{eqnarray}
\tilde{H}_{\rm 1D}^{(3)}&=&-\frac{\hbar^2}{2m}\left( \frac{\partial^2}{\partial y_1^2} + \frac{\partial^2}{\partial y_2^2} +\frac{\partial^2}{\partial y_3^2} \right) + U^{(2)}(y_{12})+ U^{(2)}(y_{23}) + U^{(2)}(y_{12}+y_{23}) \nonumber\\
&=& -\frac{\hbar^2}{m}\left(\frac{\partial^2}{\partial y_{12}^2} + \frac{\partial^2}{\partial y_{23}^2}+ \frac{\partial}{\partial y_{12}}\frac{\partial}{\partial y_{23}} \right)  - \frac{\hbar^2}{6m} \frac{\partial^2}{\partial R^2} + U^{(2)}(y_{12})+ U^{(2)}(y_{23}) + U^{(2)}(y_{12}+y_{23})\nonumber \\
&=&H^{(2)}_{\rm 1D}(y_{12})+H^{(2)}_{\rm 1D}(y_{23})+h'(y_{12},y_{23}) -\frac{\hbar^2}{6m} \frac{\partial^2}{\partial R^2}.  \label{H1d}
\end{eqnarray}
Since  the center-of-mass term can be well decoupled from the problem (or equivalently by taking zero center-of-mass momentum), $\tilde{H}_{\rm 1D}^{(3)}$ can then be reduced to Eq.(9) in the main text. Specifically,  $H_{\rm 1D}^{(2)}(y)=-\frac{\hbar^2}{m}\frac{\partial^2}{\partial y^2} +U^{(2)}(y)$ is the Hamiltonian of tetratomic system, and $h'$ follows
\begin{equation}
h'(y,y')=-\frac{\hbar^2}{m}\frac{\partial}{\partial y}\frac{\partial}{\partial y'}+U^{(2)}(y+y').
\end{equation} 
$h'$ can be viewed as the correlation between two neighboring tetratomic states, which comes from two parts: the kinetic term $(\sim \frac{\partial}{\partial y_{12}}\frac{\partial}{\partial y_{23}})$ and the long-range 1-3 interaction  ($U^{(2)}(y_{13}=y_{12}+y_{23})$). The appearance of  $\sim \frac{\partial}{\partial y_{12}}\frac{\partial}{\partial y_{23}}$ can be attributed to the non-orthogonality of $y_{12}$ and $y_{23}$. This term contributes very little to the correlation energy because of vanishing overlap between neighboring tetratomic wave-functions, while most of the correlation energy comes from the second term in $h'$, i.e., the long-range 1-3 interaction. Given the equilibrium distance between 1 and 3 as twice of $r_m$, where $r_m$ locates the minimum of $U^{(2)}$, their interaction can be approximated by the bare dipolar interaction $\sim -4C_3/(2r_m)^3$ and thus the resulted correlation energy is always negative. 

\begin{figure}[h]
	\includegraphics[width=0.85\linewidth]{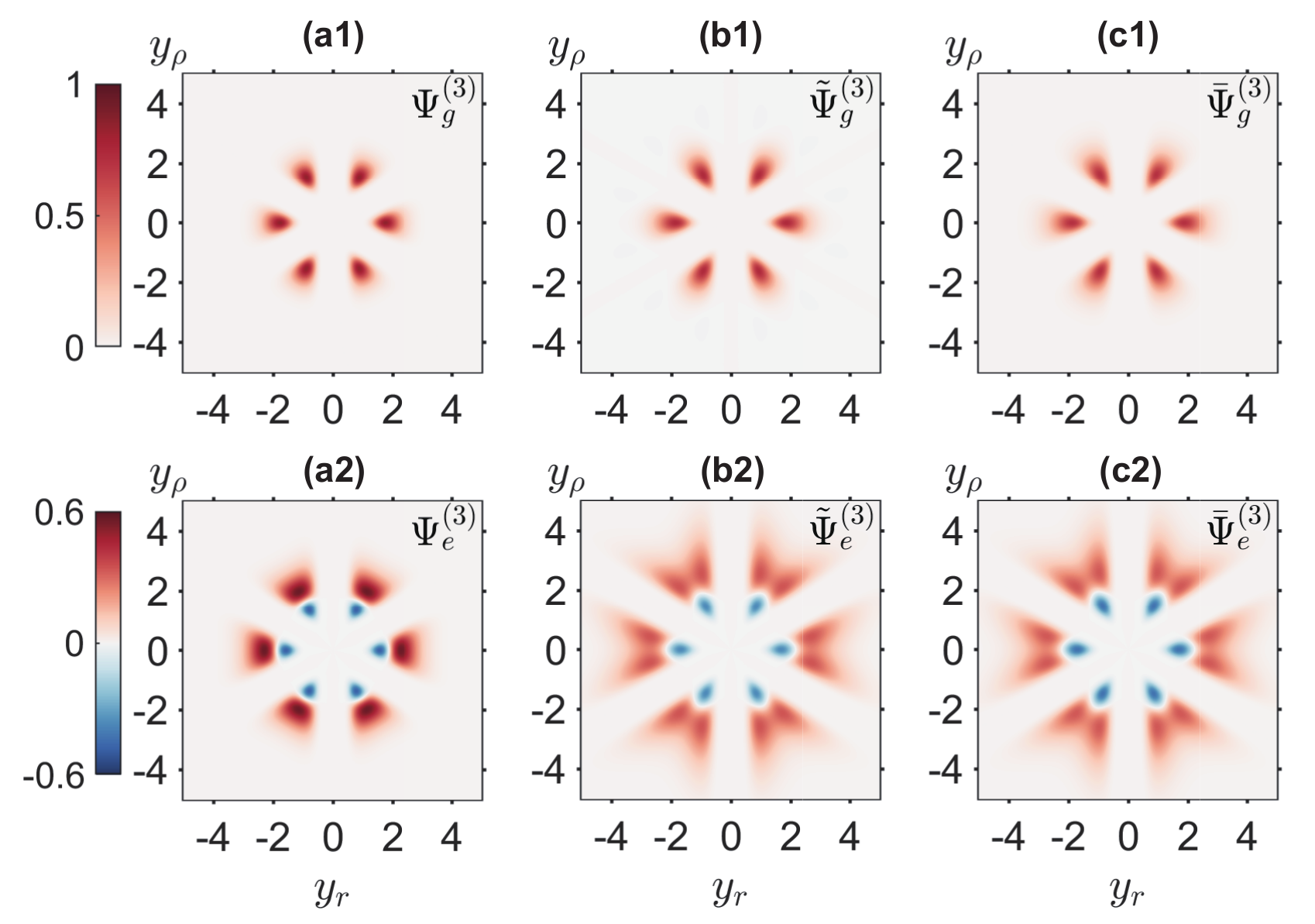}
	\caption{Hexatomic wavefunctions 
	from the full 1D model (a1,a2), in comparison with those from the construction scheme (b1,b2) and the simplified form  (\ref{wf_simplify}) (c1,c2). (a1,b1,c1) are for the ground state and (a2,b2,c2) are for the first excited state. Here we take $\hbar\Omega/E_u=52$. The length unit is $l_u$.  }  \label{fig_wf}
\end{figure}

Inspired by the decomposition in (\ref{H1d}), we can construct the hexatomic bound states from tetratomic ones. As discussed in the main text, in our construction scheme we have diagonalized a $3\times 3$ matrix expanded by $\{\Psi^{(2)}_g(y_{12})\Psi^{(2)}_g(y_{23}),\ \Psi^{(2)}_g(y_{12})\Psi^{(2)}_e(y_{23}),\ \Psi^{(2)}_e(y_{12})\Psi^{(2)}_g(y_{23})\}$, where $\Psi^{(2)}_g$ and $\Psi^{(2)}_e$ are the ground and first excited states of  tetratomic system. In Fig.\ref{fig_wf}(a1,a2), we show the hexatomic wavefunctions of two lowest  levels from the full 1D Hamiltonian $H^{(3)}_{\rm 1D}$, in comparison with (b1,b2) from the construction scheme at a given $\hbar\Omega/E_u=52$. Because of the well-separated tetratomic levels, the ground and first excited states can be  further simplified as 
\begin{eqnarray}
\bar{\Psi}^{(3)}_g&=&\Psi^{(2)}_g(y_{12})\Psi^{(2)}_g(y_{23}); \nonumber\\
\bar{\Psi}^{(3)}_e&=&\Psi^{(2)}_g(y_{12})\Psi^{(2)}_e(y_{23})+\Psi^{(2)}_e(y_{12})\Psi^{(2)}_g(y_{23}). \label{wf_simplify}
\end{eqnarray}
In Fig.\ref{fig_wf}(c1) and (c2), we also plot out these simplified $\bar{\Psi}^{(3)}_g$ and $\bar{\Psi}^{(3)}_e$, which show qualitative agreements with wavefunctions in (a1,b1) and (a2,b2).  In particular, we find excellent agreements for the ground state, where the linked tetratomic states can serve as a good approximation for  hexatomic ground state. 

Given the hexatomic ground state well approximated by $\Psi^{(2)}_g(y_{12})\Psi^{(2)}_g(y_{23})$ (for $y_{12}>0,y_{23}>0$), we now analytically evaluate the correlation function $G_2(y)$. Take $y>0$ for example, $G_2(y)$ follows
\begin{eqnarray}
\langle n(0)n(y)\rangle &=& \int dy' |\Psi^{(3)}(y_1=0,y_2=y,y_3=y')|^2 \nonumber\\
&=& \int_y^{\infty} dy' |\Psi^{(2)}_g(y)\Psi^{(2)}_g(y'-y)|^2+ \int_0^{y} dy' |\Psi^{(2)}_g(y')\Psi^{(2)}_g(y-y')|^2 +\int_{-\infty}^0 dy' |\Psi^{(2)}_g(-y')\Psi^{(2)}_g(y)|^2 \nonumber\\
&=& |\Psi^{(2)}_g(y)|^2\left( 2\int_0^{\infty} dy' |\Psi^{(2)}_g(y')|^2 \right) + \int_0^{y} dy' |\Psi^{(2)}_g(y')\Psi^{(2)}_g(y-y')|^2 \nonumber\\
&=& |\Psi^{(2)}_g(y)|^2 + \int_0^{y} dy' |\Psi^{(2)}_g(y')\Psi^{(2)}_g(y-y')|^2.
\end{eqnarray} 
In above equation, the first term gives the possibility that $y_1=0$ and $y_2=y$ stay as nearest neighbor, and the second term gives the possibility that $y_1=0$ and $y_2=y$ stay as next-nearest neighbor. This is why $G_2(y>0)$ has two peaks, as shown in Fig.3(e) of the main text, one at $\sim r_m$ and the other at $\sim 2r_m$ (here $r_m$ locates the minimum of $U^{(2)}$ or the maximum of $\Psi^{(2)}_g$).

Finally, we generalize above analysis to large ensemble of molecules and evaluate the correlation energy therein. For $N$ identical (bosonic or fermionic) molecules, the ground state under 1D description can be well approximated as  $\Psi^{(2)}_g(y_{12})\Psi^{(2)}_g(y_{23})...\Psi^{(2)}_g(y_{N-1,N})$ at given order $y_1<y_2...<y_N$. Further simplify the crystalline molecules as localized objects with equal spacing $r_m$, and approximate their interactions by the bare long-range attractive potential $\sim -4C_3/r^3$, we can estimate the correlation energy as
\begin{eqnarray}
E_{c}&\approx& \sum_{|i-j|\ge 2} U^{(2)}(y_{ij}) \nonumber\\
&=&-(N-2)\frac{4C_3}{(2r_m)^3}- (N-3)\frac{4C_3}{(3r_m)^3} -...- \frac{4C_3}{((N-1)r_m)^3} = -\frac{4C_3}{r_m^3}\sum_{i=2}^{N-1} \frac{N-i}{i^3}
\end{eqnarray} 
In the limit of $N\rightarrow\infty$, the dominant term of $E_{c}$ is given by $-\frac{4C_3}{r_m^3} N\gamma$ (with $\gamma=\sum_{i=2}^{\infty} i^{-3}$ a constant), which is negative and  linearly scales with $N$. 
Such a macroscopic correlation further lowers the energy of tetratomic chain, and thus is expected  to further stabilize such a crystalline chain along $y$ direction.

\end{document}